# ウェアラブル加速度センサからの乗車中姿勢推定の検討

A study of posture judgement on vehicles using wearable acceleration sensor

山登庸次[1]
Yoji Yamato

日本電信電話株式会社　ソフトウェアイノベーションセンタ[1]
Software Innovation Center, NTT Corporation

## 1　はじめに

IoTの適用例として，hitoe[1]等のウェアラブルセンサを用いてバイタルデータを取得し，健康や作業状態を分析して安全管理することが検討されている．センサにより加速度や心拍を取得し，そのデータをクラウド技術（例えば，[2]-[9]）を用いて管理，分析し，転倒等の姿勢や疲労蓄積を迅速に判断する．更に，Webサービス等のサービス連携技術（例えば，[10]-[21]）を用いて外部システムと連携し，スタッフ手配や緊急警告等を行う．

バスやタクシー（以後，車）のドライバーは，運転中の物拾いや疲労蓄積等は事故につながる可能性があり，ウェアラブルセンサでの安全管理にニーズがある．しかし，車内では，車加速度がウェアラブルセンサ加速度に加算されるため，正確な姿勢を推定できる保証はない．そこで，本稿ではhitoeの加速度データを用いて，乗車中の姿勢推定に関する方式を検討し，実地検証する．

## 2　既存技術

バイタルデータを取得するウェアラブルセンサとして，多彩な端末が出てきている．Apple Watchは，腕時計型で，心拍，加速度センサを内蔵している．hitoeはNTTと東レが開発したTシャツ型センサで，心電図と3軸加速度が，Tシャツを着用するだけで取得できる（前方がY軸，頭部方向がZ軸）．

これらセンサで取得される加速度を分析し，姿勢を推定するSDKも市中に出ている．しかし，乗車中は，車の加速度が加わり，正確な姿勢を推定できるか不明であるため，乗車中の姿勢推定の検証が必要である．

## 3　方式検討

乗車中の姿勢推定には2点考慮する点がある．第一に，ウェアラブルセンサには車の加速度が加算されて測定される．第二に，ドライバーの安全管理を想定した際に，運転中に席を離れることは想定外であるため，運転中の物拾いやよそ見等が検出したい危険姿勢である．

上記を踏まえ，まず，車の加速度を減算することで，乗車ドライバーの姿勢を推定する方法が考えられる（方式1：hitoe加速度からスマホ加速度を減算）．ドライバーが着るhitoe加速度から，ドライバーが保持するスマホの加速度を，タイムスタンプを合わせた後，減算する形である．ドライバー自体の加速度が抽出できるが，2つのセンサの精度違い等減算の実現性が懸念される．

次に，物拾い等検出したい特定姿勢の加速度パターンを，hitoe加速度データから分析する方法が考えられる（方式2：hitoe加速度パターンから特定姿勢を分析）．バ

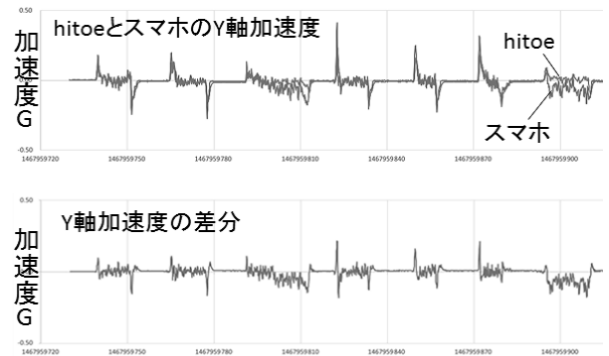

図1　台車移動中のhitoeとスマホのY軸加速度データ

スやタクシーは急発進を避けるため，X，Y軸の加速度はそれほど大きくない．一方，物拾い等は前屈等の姿勢変化があるため，重力の影響でY, Z軸の加速度が大きく変わる．そこで，加速度変化に着目すれば，閾値等の設定で，特定の危険姿勢を判定できる可能性がある．

## 4　方式検証

方式1について，hitoeとスマホ（Xperia）が同じ動きの場合に，加速度を減算してゼロにできるかを確認するため，hitoeとスマホを台車に乗せ，発進，直進，右左折等を行った．図1にその際のY軸のデータを示す．図1より，hitoeとスマホの加速度データは類似だが，減算してもゼロにはならず，ローパスフィルタを通して高周波ノイズを除去した場合も完全にゼロにはならない．これは，2つのセンサが，反応や精度等が異なるためと考えられ，方式1には課題が多い．

方式2について，乗車中の姿勢変更がhitoe加速度データから検出できるかを確認するため，hitoe着用者が路線バスに乗車し，路線バスで着席中に物を拾う等の姿勢変更を行った．図2にその際のデータを示す．図2より，バス起因の加速度変化は，それほど大きくなく，また，X, Y軸の細かい加速度変化であることが分かる．一方，着席中の物拾いは，体の傾きが変わるため，シャツの前方であるY軸の加速度変化が大きいことが分かる．そのため，Y軸の加速度変化に着目することで，物拾い等の姿勢を検出可能である．具体的には，閾値20度（-0.34G）とすることで，路線バスでは検出可能であった．

ただし，Y軸の加速度変化を閾値にした場合，急傾斜な路線の際は，誤判定する懸念がある．そこで，急勾配（20%程度）で知られる，富士山登坂バスでも検証を行った．その結果，坂道であっても，閾値20度までの加速度変化には至らなかったため，誤判定は生じなかった．

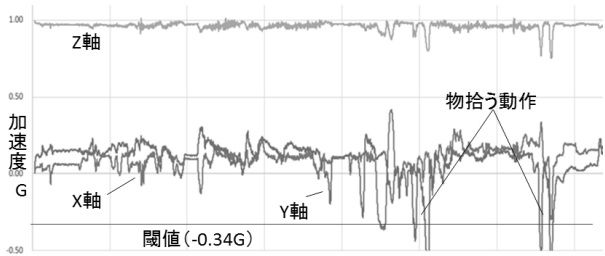

図 2 路線バス（平地）中の hitoe 加速度データ